# Standards for Graph Algorithm Primitives

Tim Mattson (Intel Corporation), David Bader (Georgia Institute of Technology), Jon Berry (Sandia National Laboratory), Aydin Buluc (Lawrence Berkeley National Laboratory), Jack Dongarra (University of Tennessee), Christos Faloutsos (Carnegie Melon University), John Feo (Pacific Northwest National Laboratory), John Gilbert (University of California at Santa Barbara), Joseph Gonzalez (University of California at Berkeley), Bruce Hendrickson (Sandia National Laboratory), Jeremy Kepner (Massachusetts Institute of Technology), Charles Leiserson (Massachusetts Institute of Technology), Andrew Lumsdaine (Indiana University), David Padua (University of Illinois at Urbana-Champaign), Stephen Poole (Oak Ridge National Laboratory), Steve Reinhardt (Cray Corporation), Mike Stonebraker (Massachusetts Institute of Technology), Steve Wallach (Convey Corporation), Andrew Yoo (Lawrence Livermore National Laboratory)

*Abstract*— It is our view that the state of the art in constructing a large collection of graph algorithms in terms of linear algebraic operations is mature enough to support the emergence of a standard set of primitive building blocks. This paper is a position paper defining the problem and announcing our intention to launch an open effort to define this standard.

*Keywords-component; Graphs; Algorithms; Linear Algebra; Software Standards*

## I. PROBLEM STATEMENT

Data analytics and the closely related field of "big data" have emerged as a leading research topics in both applied and theoretical computer science. While it has been shown that many problems can be addressed with a "map-reduce" style framework, as we move to the next level of sophistication in data analytics applications, graph algorithms that demand more than "map-reduce" will play an increasingly vital role. There are many ways to organize a collection of graph algorithms into a high level library to support data analytics. It is probably premature to standardize these graph APIs. The low level building blocks of graph algorithms, however, are well understood and we believe a suitable target for standardization. In particular, the representation of graphs as sparse matrices allows many graph algorithms to be represented in terms of a modest set of linear algebra operations [1,2,5].

Our concern, however, is that as new researchers enter this expanding field of research, the linear algebraic foundation of this class of graph algorithms will fragment. Diversity at the level of the primitive building blocks of graph algorithms will not help advance the field of graph algorithms. It will hinder progress as groups create different overlapping variants of what should be common low level building blocks. Furthermore, diverse sets of primitives will complicate the ability of the vendor community to support this research with math tuned to the needs of these algorithms.

It is our view that the state of the art in constructing a large collection of graph algorithms in terms of linear algebraic operations is mature enough to support the emergence of a standard set of primitive building blocks. We believe it is critical that we move quickly so as new research groups enter this field we can prevent needless and ultimately damaging diversity at the level of the basic primitives supporting this research; thereby freeing up researchers to innovate and diversify at the level of higher level algorithms and graph analytics applications.

## II. THE STATE-OF-THE-ART

The standardization of sparse linear algebra historically begins with the NIST Sparse Basic Linear Algebra Subprograms (BLAS) [3] and consists of Sparse Vector (Level 1), Matrix Vector (Level 2), and Matrix Matrix (Level 3) operations. These BLAS were designed for solving the kinds of sparse linear algebra operations that arise in finite element simulation techniques that are widely used in engineering. In particular, the operations are limited to traditional multiplication and addition operations and, in the case of matrix-matrix multiply, usually one of the arguments is dense.

We can extend the BLAS to address the needs of graph algorithms by generalizing the pair of operations involved in the computations to define a semiring. For example, in semiring notation we could write the most common operations found in the exiting Sparse BLAS as

$$C = A +.* B$$

where +.* denotes standard matrix multiply. In the case of the NIST Sparse BLAS, **A** is a sparse matrix, and **B** and **C** are usually tall skinny dense matrices. In graph algorithms, a fundamental operation is matrix-matrix multiply where both matrices are sparse. This operation represents multi source 1-hop breadth first search (BFS) and combine, which is the foundation of many graph algorithms. In addition, it is often the case that operations other than standard matrix multiply are desired, for example:

$$C = A \text{ max.}+ B$$

$$C = A \text{ min.max } B$$

$$C = A \mathbin{|.\&} B$$

$$C = A \, f().g() \, B$$

With this more general case of sparse matrix multiply, a wide range of graphs algorithms can be implemented [1]. An implementation of this approach is found in the combinatorial BLAS [2]. The combinatorial BLAS provides implementations of nine functions: matrix-matrix multiply (multi source BFS combine), matrix-vector multiply (single source BFS combine), element wise matrix-matrix multiply (edge weighting), reduce (in/out degree), sub reference (sub-graph selection), sub assign (sub-graph insertion), scale matrix (edge weighting), scale vector (vertex weighting), and apply unary operator (edge transformation). These functions are a reasonable basis for graph algorithms primitives, and the combinatorial BLAS demonstrate that they can be effectively implemented in a parallel.

The data types and storage formats are also an important consideration. The values should be able to handle all the standard types: float, double, complex, boolean, signed/unsigned integers of various lengths, and pointers to external user defined data structures. There should be a concept of a symmetric matrix to handle undirected graphs. Internal formats should include at least Compressed Sparse Rows (CSR) and Compressed Sparse Columns (CSC). Other formats such as tuples and more complex formats [4] should also be considered.

## III. RECOMMENDATIONS

We believe the research community working on graph algorithms expressed as linear algebra should come together now and define a common API they can use in their research. The combinatorial BLAS are an excellent starting point for this work. We propose the formation of an ongoing combinatorial BLAS forum to finalize the specification and take ownership of its ongoing evolution.